\documentclass{PoS2}

\title{Measurement of Z/$\gamma^*$+jet+X and $\gamma$+b/c+X Cross Sections with the D0 Detector}

\ShortTitle{Measurement of Z/$\gamma$+jet+X and $\gamma$+b/c+X Cross Sections with the D0 Detector}

\author{\speaker{Murilo RANGEL}\\
        Laboratoire de l'Acc\'el\'erateur Lin\'eaire\\
        E-mail: \email{rangel@fnal.gov}}
\author{on behalf of D0 collaboration} 

\abstract{
We present measurements of differential cross sections for the inclusive $Z/\gamma^*$ +
jet production and the inclusive photon plus heavy flavor production
in a data sample of 1fb$^{-1}$ collected with the D\O\ detector in proton-antiproton
collisions at $\sqrt{s}$=1.96 TeV.
In the first measurement, we compare kinematic distributions of the
$Z/\gamma^*$ and the jets as well as various angles of the $Z$+jet system
with different Monte Carlo event generators and next-to-leading order
perturbative QCD (NLO pQCD) predictions with non-perturbative corrections
applied.
In the second measurement, we compare the results with NLO pQCD
predictions, covering photon transverse momenta 30-150 GeV,
photon rapidities $|y^{\gamma}| < 1.0$, jet rapidities $|y^{jet}| < 0.8$,
and jet transverse momenta $p_{T}^{jet} > 15$ GeV.
}

\FullConference{European Physical Society Europhysics Conference on High Energy Physics\\
		 July 16-22, 2009\\
		 Krakow, Poland}

\begin{document}

\section{Introduction}

Understanding the background productions in searches for new physics is
very challenging at hadron colliders. Therefore, measurements of 
differential cross sections represent important milestones in the
discovery road. In this note, we present measurements of $Z/\gamma^*$ +
jets production and photon plus heavy flavor jets at the Fermilab Tevatron
with the D\O\ detector \cite{d0detector}.

The production of vector bosons is an important signal at hadron colliders,
providing unique information about the production mechanism of heavy bosons
with additional hard partons. The electron and muon decay
modes are distinct experimental signatures, and can be identified with low
background rates. 

Photons produced in association with heavy quarks (c or b)
at hadron colliders
provide valuable information about the heavy quark and
gluon content of the initial state hadrons, .
Although the background rates are not negligible, the statistics are
high enough to constrain accurately the parton distribution
functions (PDFs).

\section{$Z/\gamma^{*}$+jets}

The D\O\ collaboration has studied the production of $Z/\gamma^*$ +
jets in both the electron channel and the muon channel \cite{zjets1, zjets2, zjets3}.
The jets were reconstructed using a seeded mid-point cone algorithm
\cite{d0jets} with cone size of 0.5, and they are required to
have $|y| < $ 2.8 and $p_T > $ 20 GeV.

\begin{figure}[ht]
\begin{center}
\includegraphics[width=0.25\textwidth]{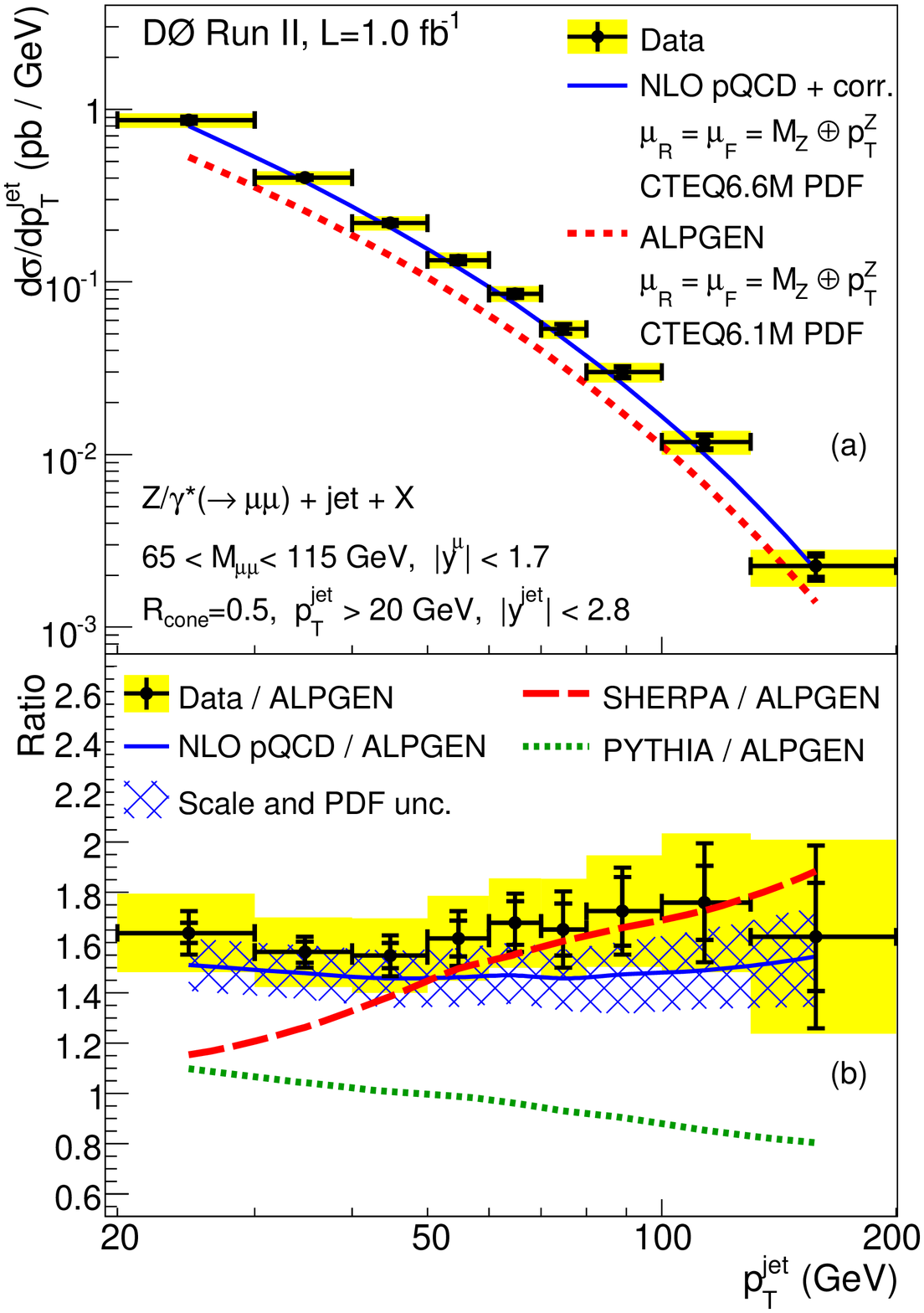}
\includegraphics[width=0.25\textwidth]{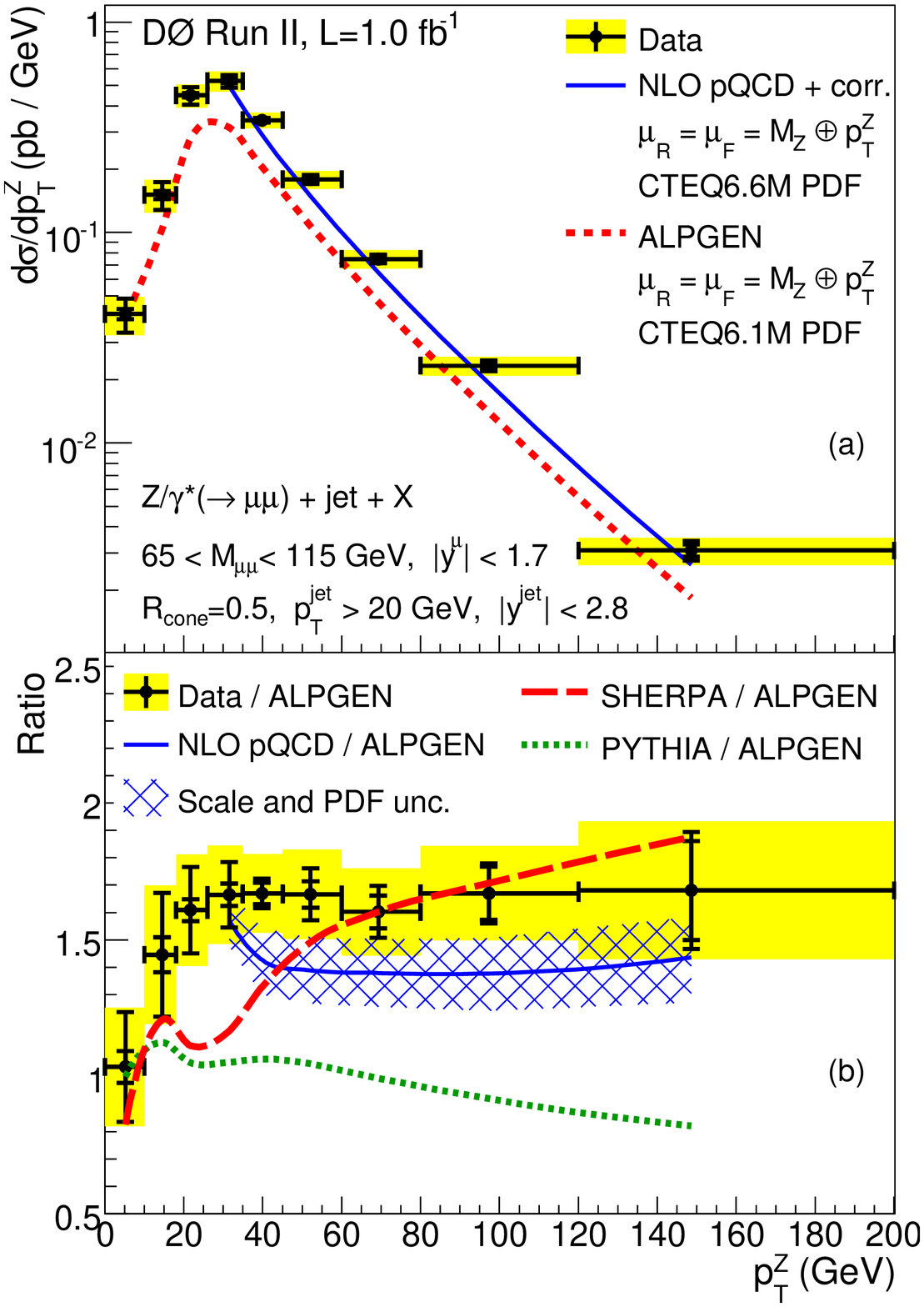}
\caption{
The measured cross section (a) and the ratio of
data and predictions to ALPGEN (b) are shown in bins of leading
jet $p_T$ (left) and in bins of $Z/\gamma^* p_T$
(right).
\label{fig:z1}}
\end{center}
\end{figure}

The muons were selected to have opposite charge, $p_T > $ 15 GeV, $|\eta| < $ 1.7
and the di-muon invariant mass ranging between 65 GeV and 115 GeV.
Isolation requirements were used to reduce the background rates to negligible
levels.
The electrons were selected to have opposite charge, $p_T > $ 25 GeV, $|\eta| < $ 1.1
or 1.5 $< |\eta| <$ 2.5,
and the di-electron invariant mass ranging between 65 GeV and 115 GeV.
In the muon decay channel, we measured differential cross section in the 
leading (transverse momenta) jet $p_T$ and the $Z/\gamma^* p_T$
\cite{zjets1}. In Figure \ref{fig:z1}, the
NLO pQCD predictions 
and from three event generators (ALPGEN \cite{alpgen}, using PYTHIA \cite{pythia}
for the showering; SHERPA \cite{sherpa};
PYTHIA, with all jets coming from the parton shower)
are compared to data.

\begin{figure}[ht]
\begin{center}
\includegraphics[width=0.49\textwidth]{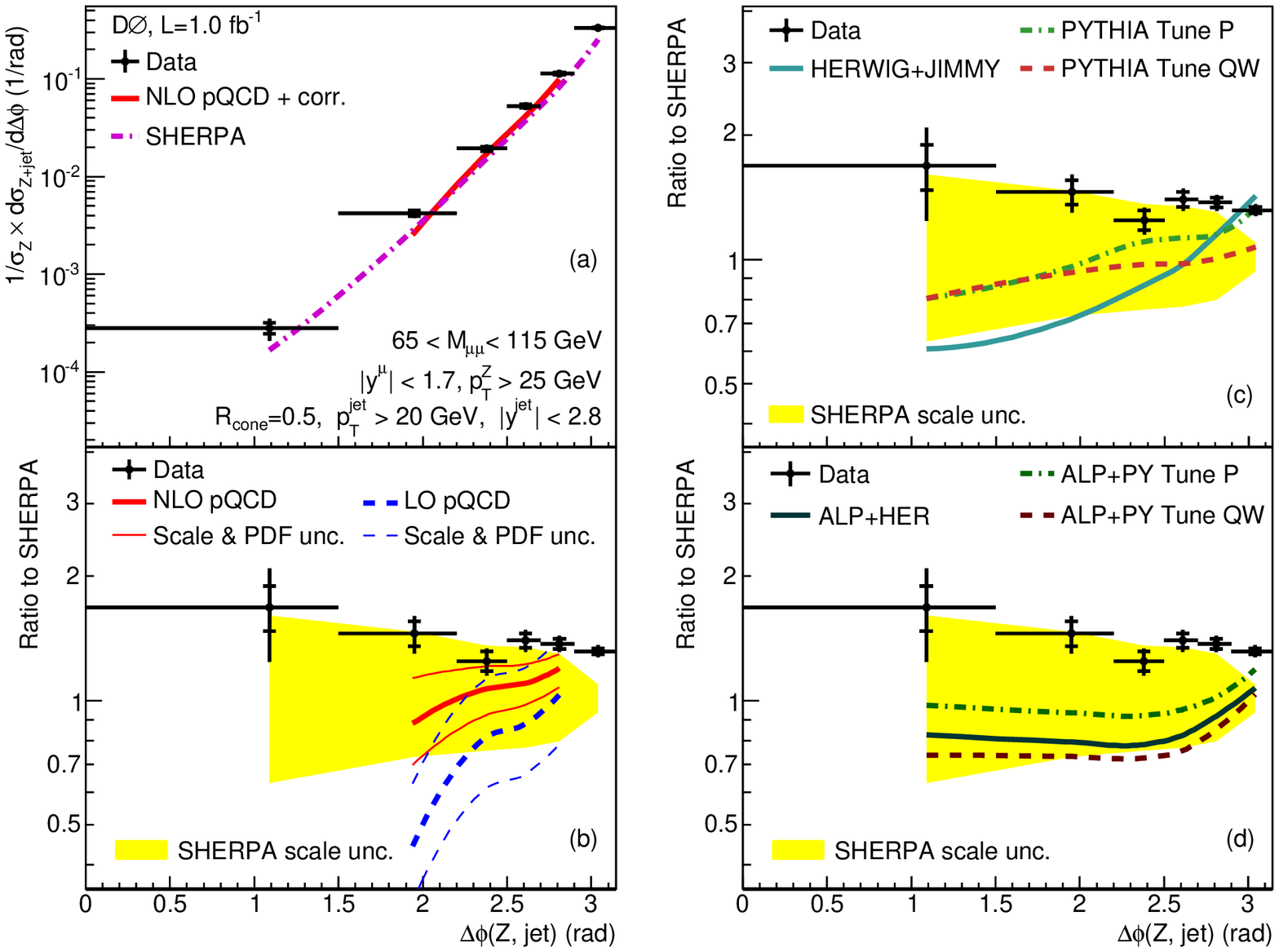}
\includegraphics[width=0.49\textwidth]{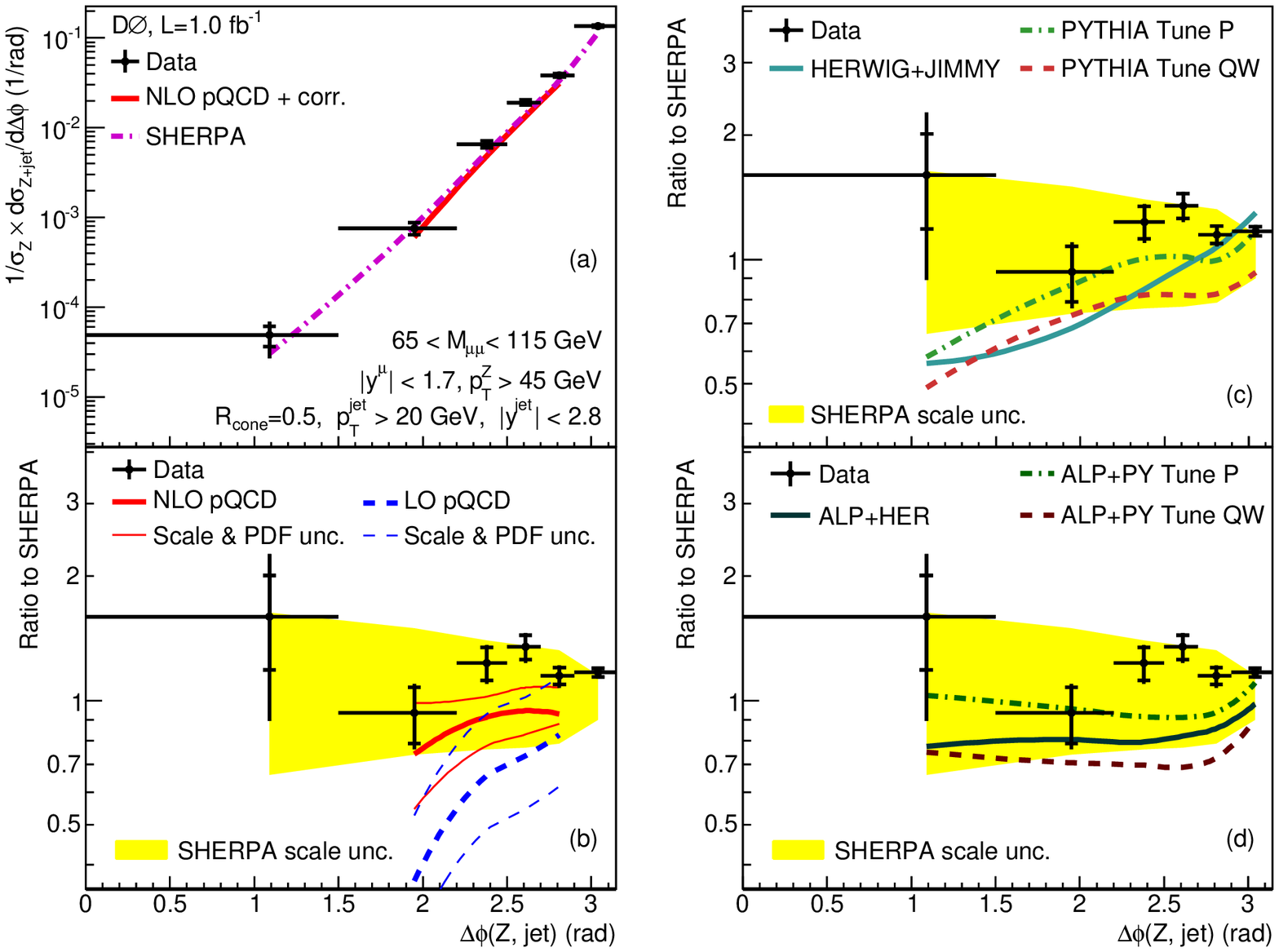}
\caption{
The measured normalized cross section in bins of $\Delta \varphi (Z, jet)$ for
$Z p_T >$ 25 GeV (left) and for $p_T^{Z} >$ 45 GeV (right).
\label{fig:angle}}
\end{center}
\end{figure}

The $\Delta \varphi$ between the $Z/\gamma^*$ and the leading jet were
also measured in the muon channel using two different $p_T$ thresholds (25 GeV and 45 GeV),
since the production of additional jets is essentially uncorrelated with the 
$Z/\gamma^*$ production \cite{zjets3}. We compared the measurement with
NLO pQCD and LO pQCD predictions, and the event generators: 
SHERPA; HERWIG using JIMMY \cite{JIMMY} for multiple parton interactions;
PYTHIA with tune QW and with the Perugia tune using the $p_T$
ordered shower; ALPGEN,
using the mentioned PYTHIA tunes and HERWIG for the showering
(Figure \ref{fig:angle}).

\begin{figure}[ht]
\begin{center}
\includegraphics[width=0.49\textwidth]{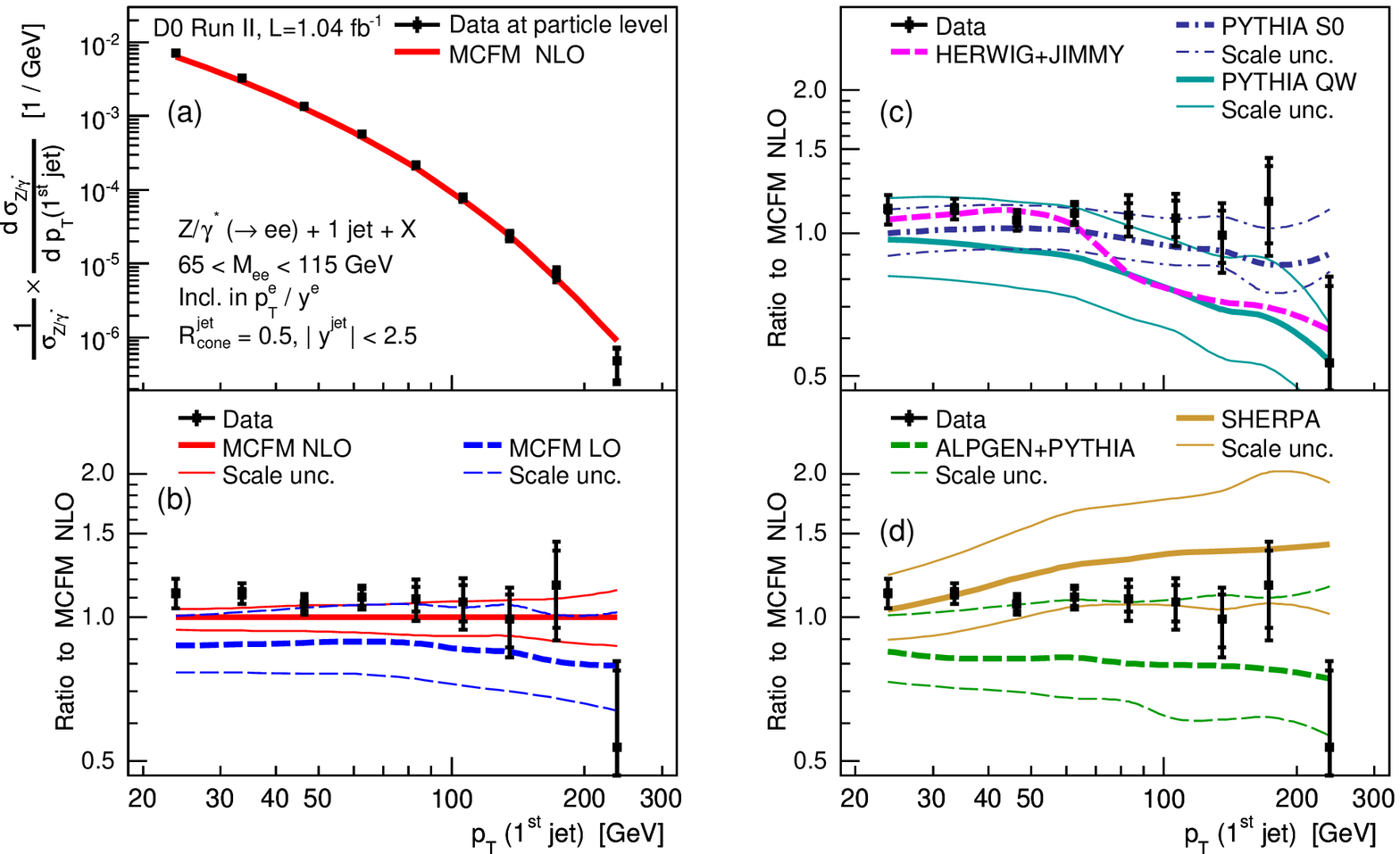}
\includegraphics[width=0.49\textwidth]{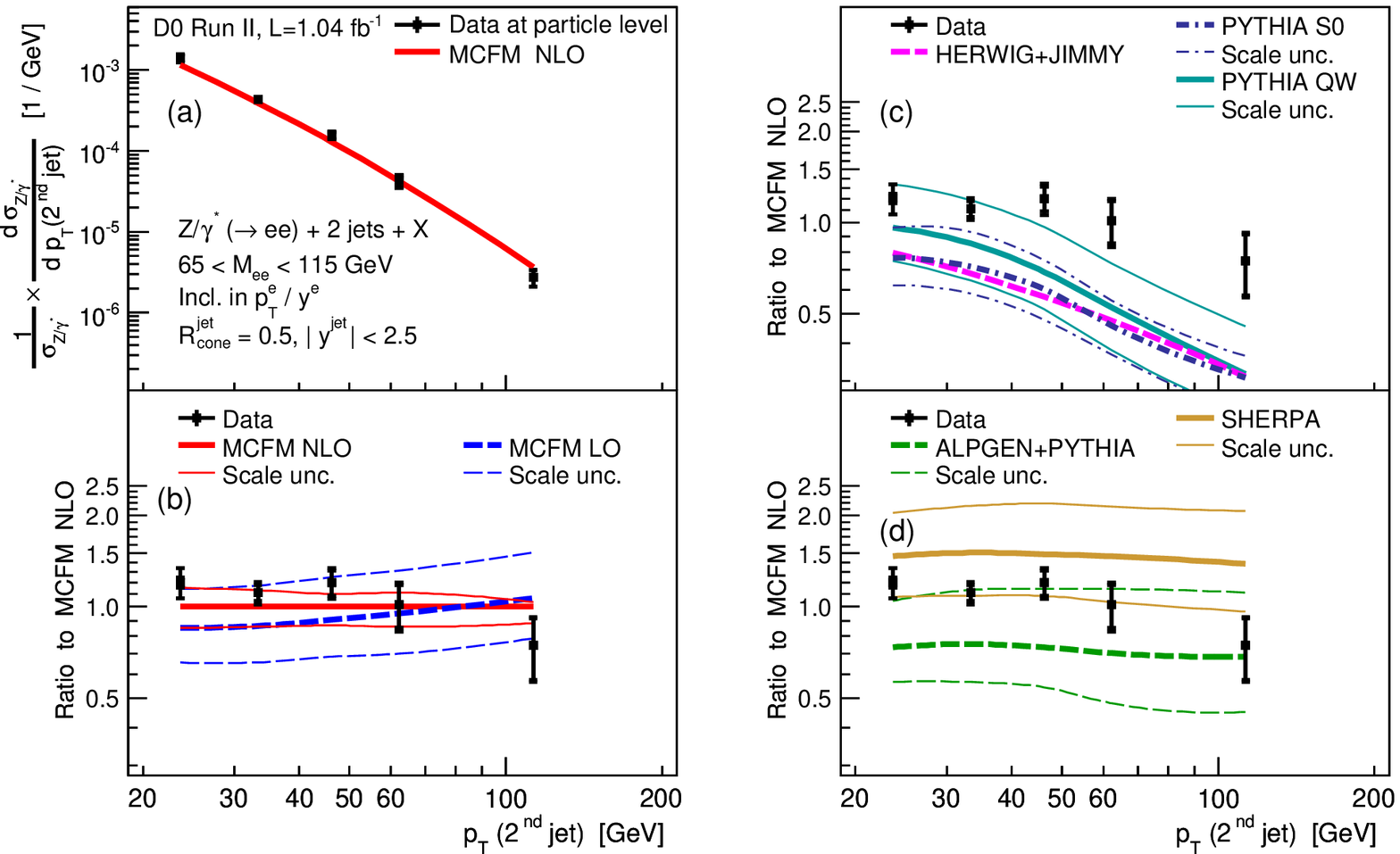}
\includegraphics[width=0.49\textwidth]{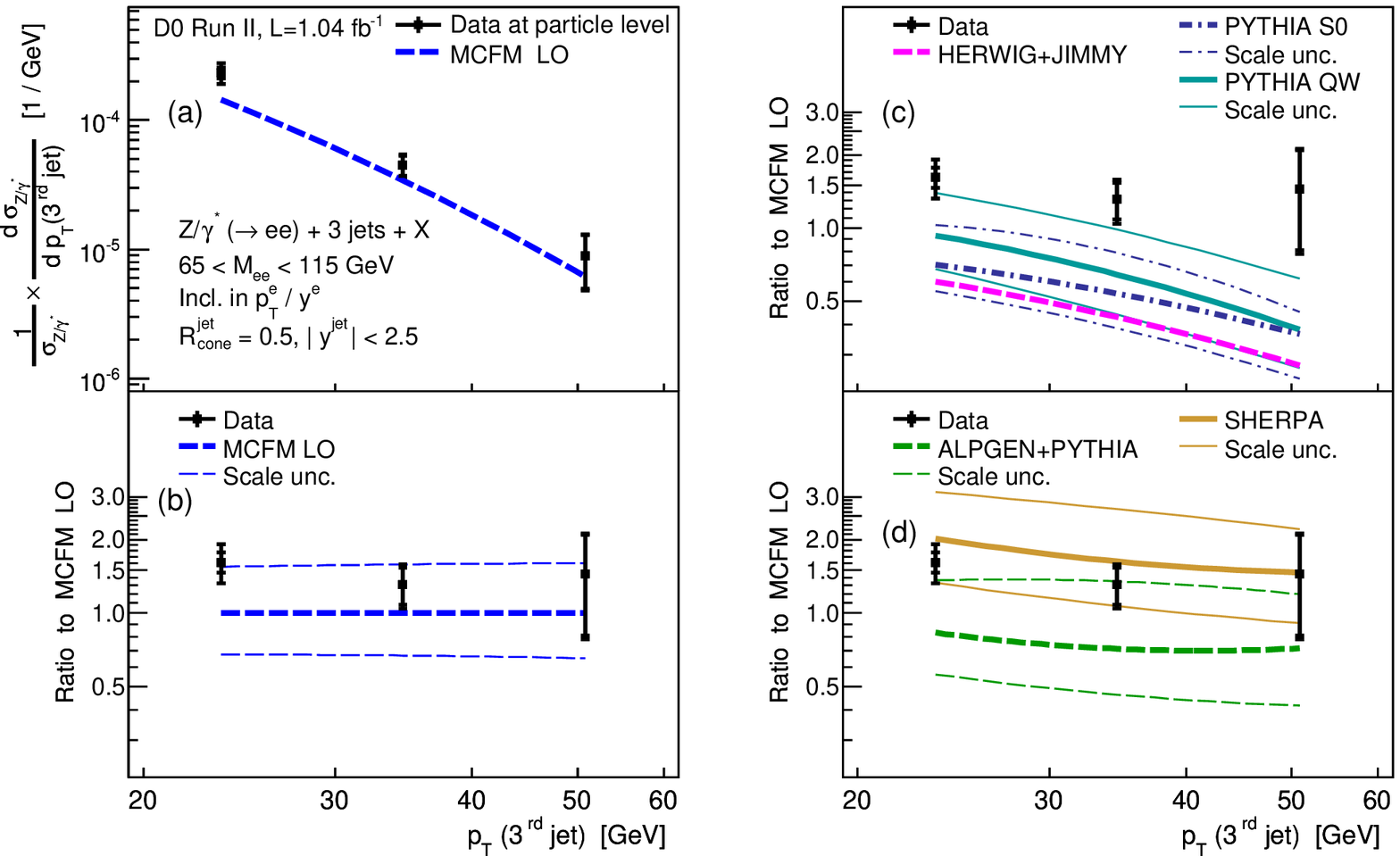}
\caption{
The measured normalized cross section for the leading jet in $Z/\gamma^*+jet+X$ events
(upper left) compared to the predictions of NLO and the ratios of data and theory predictions to NLO
compared to various event generator models.
The measured normalized cross section for the second leading jet in $Z/\gamma^*+2 jets+X$ events
(upper right) compared to the predictions of NLO and the ratios of data and theory predictions to NLO
compared to various event generator models.
The measured normalized cross section for the third leading jet in $Z/\gamma^*+3 jets+X$ events
(bottom) compared to the predictions of LO and the ratios of data and theory predictions to LO
compared to various event generator models.
\label{fig:3jets}}
\end{center}
\end{figure}

In the electron decay channel, we measured the jet $p_T$
spectra normalized to the $Z/\gamma^*(\rightarrow e^+e^-)+X$ cross section
in different jet multiplicities \cite{zjets2}.
The measurements were compared to different theory predictions:
NLO pQCD; LO pQCD; PYTHIA using tune QW; PYTHIA using Tune S0;
HERWIG using JIMMY; ALPGEN using PYTHIA tune QW; and SHERPA
(Figure \ref{fig:3jets}). 

The pQCD NLO prediction describes the D\O\ measurements within uncertainties
while the event generators show varying agreement. With more data,
these results can be extended and tighter constraints can be placed.

\section{Photon+heavy flavor}

Using the D\O\ detector, we studied events with at least one photon candidate
and at least one heavy-flavor jet candidate \cite{photonhf}.
The photons were selected to have $p_T >$ 30 GeV with $|y| <$ 1.0 and
the leading jet  $p_T$ > 15 GeV and $|y| <$ 0.8.
To suppress background events coming from cosmic-ray muons and $W$ leptonic decays,
the total missing transverse energy was required to be less than 70\% of the
photon $p_T$.
The remaining background from dijet events,
containing $\pi^{0}$ and $\eta$ mesons that can mimic photon signatures, is
rejected using an artificial neural network (ANN) with the
requirement that the ANN output be > 0.7.
Light jets are suppressed using another dedicated ANN (b-ANN),
trained to discriminate light flavor from heavy flavor jets.
The leading jet is required to have a b-ANN output value > 0.85.


The fraction of c and b jets in the final data sample is determined using a fitting
technique, where the discriminant is $P_b = - \ln \prod_i P_i$
, where $P_i$ is the probability of a track in the jet cone
to originate from the primary vertex, omitting the least likely track
to have come from this vertex.

\begin{figure}[ht]
\begin{center}
\includegraphics[width=0.4\textwidth]{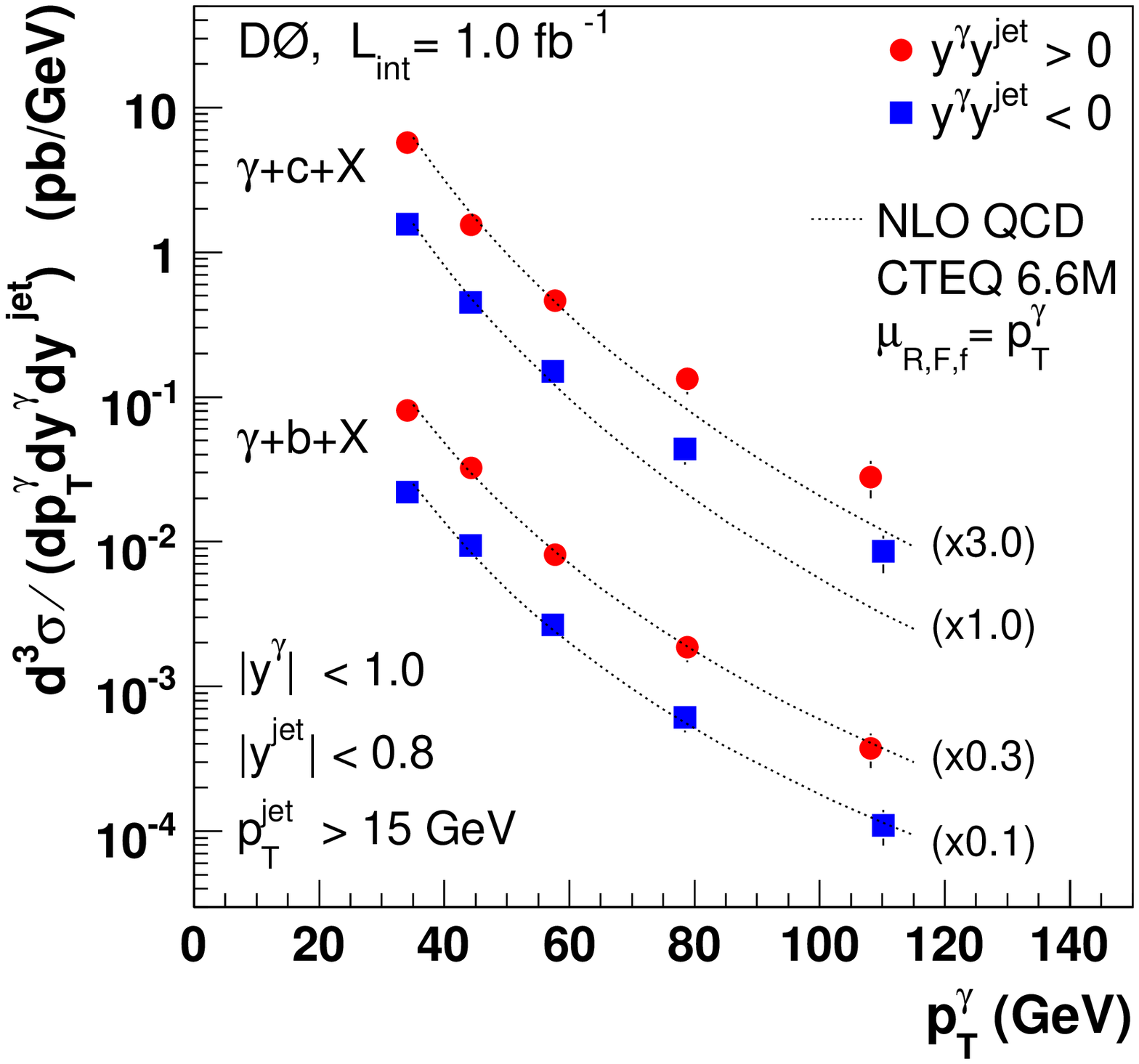}
\includegraphics[width=0.4\textwidth]{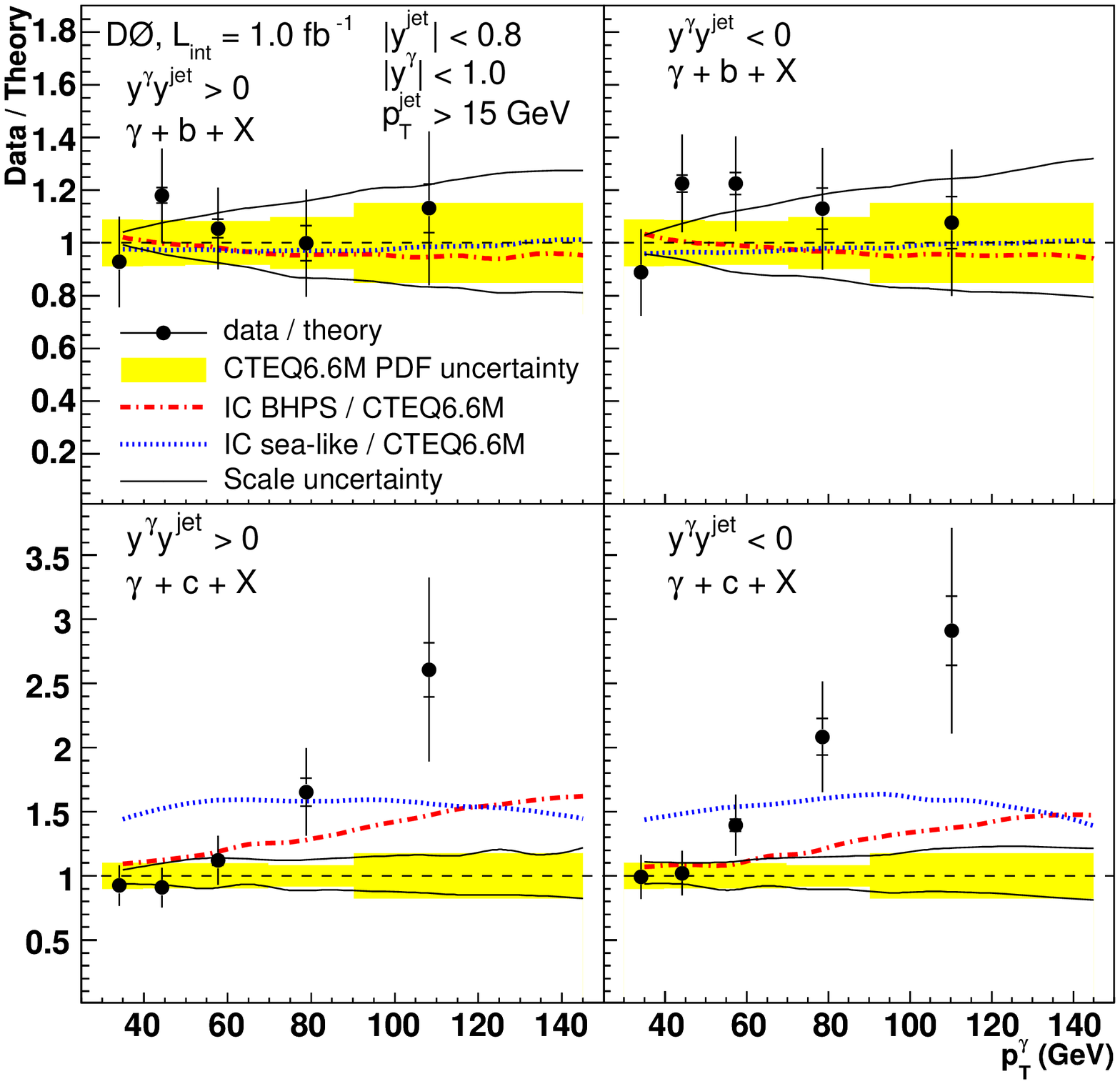}
\caption{The $\gamma + b$ and $\gamma + c$ differential cross sections as a function of
$p_T^{\gamma}$ for both rapidity
regions. The data points include the overall uncertainties from the measurement, and the
theoretical predictions are displayed as dotted lines. The uncertainties from the
theoretical predictions include those from the CTEQ 6.6M PDFs (yellow band) and from the
choice of scale (full line). The ratio of two intrinsic charm models to the standard theoretical
predictions are also included (dashed lines).
\label{fig:photonhf}}
\end{center}
\end{figure}

The measured differential cross sections and their ratios to
theoretical predictions are presented in five bins of
$p_T^{\gamma}$ and two regions of $y^{\gamma}y^{jet}$ ($y^{\gamma}y^{jet} > 0$ 
and $y^{\gamma}y^{jet} < 0$),
and can be seen in Figure \ref{fig:photonhf} for photon + b jets and 
photon + c jets. Theoretical predictions are from  
NLO pQCD calculations using the CTEQ 6.6M PDFs.
For photon + c jets, comparisons with CTEQ 6.6M PDFs based on the models
with an intrinsic charm component (IC) were also done.

The NLO pQCD prediction agrees with the measured cross
sections for photon+b production over the entire $p_T^{\gamma}$ range,
and with photon+c production for $p_T^{\gamma}$ < 70 GeV.
For $p_T^{\gamma}$  > 70 GeV, the measured photon + c cross section is higher
than the NLO pQCD prediction by about 1.6 - 2.2 standard deviations
(including only the experimental uncertainties) with the
difference increasing with growing  $p_T^{\gamma}$.


\section{Conclusion}

Important measurements have been performed with the D\O\ detector,
testing NLO pQCD, and the modeling of these complex final
states by event generators. The understanding of the discrepancies observed between data 
and predictions is vital to the sensitivity to new physics at the Tevatron
and LHC.

\end{document}